\documentclass[preprint,showpacs,preprintnumbers,amsmath,amssymb]{revtex4}

\usepackage{graphicx}
\usepackage{dcolumn}
\usepackage{bm}

\def\slash#1{\not\!#1}

\def\delsla{\!\!\not\!\partial}

\begin{document}
\preprint{OCU-293}

\title{Phase diagram of quark-antiquark and diquark condensates\\
 in the 3-dimensional Gross Neveu model\\
 with the 4-component spinor representation}

\author{Hiroaki Kohyama}
 \email{kohyama@sci.osaka-cu.ac.jp}
\affiliation{%
Department of Physics, Osaka City University, 
Sumiyoshi-ku, Osaka 558-8585, JAPAN
}%

\date{\today}
\begin{abstract}
We construct the phase diagram of the quark-antiquark and diquark condensates
at finite temperature and density in the $2 + 1$ dimensional (3D) two flavor
massless Gross Neveu (GN) model with the 4-component quarks. 
In contrast to the case of the 2-component quarks,
there appears the coexisting phase of the quark-antiquark and diquark condensates.
This is the crucial difference between the 2-component and 4-component quark cases
in the 3D GN model.
The coexisting phase is also seen in the 4D Nambu Jona-Lasinio model.
Then we see that the 3D GN model with the 4-component quarks
bears closer resemblance to the 4D Nambu Jona-Lasinio model.
\end{abstract}

\maketitle

\section{\label{sec:level1}Introduction}
In recent years, a lot of works has been devoted to the study of the phase
structure of Quantum Chromodynamics (QCD). QCD is an asymptotically free theory
and the interactions between quarks and gluons become weak at high energy\cite{Gross}.
Then, at high temperature and/or density, the quarks and gluons constitute rather
weakly interacting system, which is called the quark-gluon plasma. On the other
hand, at low temperature and density, quarks and gluons are confined into hadrons
and can not be observed as free particles. Furthermore, the existence of the
color superconducting phases at low temperature and moderate baryon density
has been widely accepted. The color superconductivity is the state where the
quark-quark (diquark) Cooper pairs\cite{Cooper} are induced by the attractive
interaction in the color antitriplet channel\cite{Wilczek}.

One of the simple theories to study the above mentioned subject is the
Nambu Jona-Lasinio (NJL) model, which is a low energy effective field theory
of QCD\cite{NJL}. The NJL model successfully describes the QCD phase structure,
and a variety of works has been devoted to the study on the basis of the NJL model.
(For nice reviews, see, e.g. \cite{HatsuKuni,Mei}.) In particular, through analysis
on the competition between quark-antiquark ($q\bar{q}$) and diquark ($qq$) condensates
in the NJL model, it has been found that there appears the coexisting phase of the
$q\bar{q}$ and $qq$ condensed phases (see \cite{Mei}).

The study of the phase structure of the NJL model in lower dimensions (D)
is also an interesting issue since the models usually become simpler in lower 
dimensions\cite{Weinberg}. Indeed, the Gross Neveu (GN) model\cite{GN}, which is the
counterpart of the NJL model in 2D and 3D, becomes renormalizable.
Recently, the phase structure of the
$q\bar{q}$ and $qq$ condensates was studied within the 3D GN model\cite{Kohyama,Niegawa}.
There the quarks are assigned to the lowest nontrivial
(2-dimensional) representation of the $O(2,1)$ group which we refer to as
the 2-component (2c) quarks. The resultant phase diagrams show that there does not appear
the region where the $q\bar{q}$ and $qq$ condensates coexist, which is in sharp contrast
to the 4D NJL case. This difference may stem from the difference in form of
the $qq$ condensate term in the Lagrangian density between both cases.
In the 2c quark case in 3D, the $qq$ condensate term in the Lagrangian does
not include the $\gamma^5$ matrix, which leads to the above-mentioned difference.
However, as discussed in \cite{Appel}, the $\gamma^5$ enters in the case of the 4c quarks in 3D.
In this sense, the 3D GN model with the 4c quarks is expected to bear closer resemblance
to the 4D NJL model. Actually, in vacuum (zero temperature and chemical potential) theory,
the coexisting phase of the $q\bar{q}$ and $qq$ condensates
has been found  in the 3D GN model with the 4c quarks\cite{Klimenko}.

In this paper, we study the $q\bar{q}$ and $qq$ condensates at finite temperature
and density in the 3D 2 flavor massless GN model with the 4c quarks.
We construct the phase diagrams, and discuss the similarities and differences among 
the models, the 3D GN model with the 4c quarks, the 3D GN model with the 2c quarks
and the 4D NJL model.

The paper is organized as follows: In Sec.II we introduce the Lagrangian density
in the 3D GN model with the 4c quarks and employ the mean-field
approximation. In Sec.III the derivation of the thermodynamic potential is given.
Then, in Sec.IV, we show the numerical results for the $q\bar{q}$ and $qq$ condensates 
at zero and finite temperature. In Sec.V, the phase diagrams of the
$q\bar{q}$ and $qq$ condensates are obtained. Sec.VI is devoted to concluding
remarks.

\section{3D Gross Neveu Model}

Following the same reasoning as described in \cite{Mei}, we employ the Lagrangian
density:
\begin{align}                                                                 
\mathcal{L} = & \bar{q}i \delsla q + G_S (\bar{q}q)^2 \nonumber \\
  &+ G_D(\bar{q} i \tau_2 \lambda_2 \gamma_5 q^C) 
        (\bar{q}^C i \tau_2 \lambda_2 \gamma_5 q)
\label{GN4C}.
\end{align}                                                                   
The notations used in Eq.(\ref{GN4C}) are the same as in the 3D 
2 flavor massless GN model
with the 2c quarks given in \cite{Kohyama}, except that the
$\gamma$ matrices are $4 \times 4$ matrices here.
For the $\gamma$ matrices, we use the same form as in \cite{Appel},
\begin{align}                                                                 
&\gamma^0 = 
\left(
\begin{array}{cc}
\tau_3 & 0\\
0 & -\tau_3
\end{array}\right)\!,\,
\gamma^1 = 
\left(
\begin{array}{cc}
i\tau_1 & 0\\
0 & -i\tau_1
\end{array}\right)\!, \nonumber \\
&\gamma^2 = 
\left(
\begin{array}{cc}
i\tau_2 & 0\\
0 & -i\tau_2
\end{array}\right)\!,\,
\gamma^5 = i
\left(
\begin{array}{cc}
0 & 1\\
-1 & 0
\end{array}\right)\!.
\end{align}                                                                   
The charge conjugated fields are given by $q^C = C\bar{q}^T, \, \, \bar{q}^C=q^T C$,
where $C = \gamma^2$.

Employing the mean-field approximation, we have
\begin{align}                                                                 
\tilde{\mathcal{L}} = & \bar{q} (i \delsla
     - \sigma) q 
    + \frac{1}{2}\Delta^{*}(\bar{q}^C i\tau_2 \lambda_2 \gamma_5 q) \nonumber \\
    &{} + \frac{1}{2}\Delta(\bar{q} i\tau_2 \lambda_2 \gamma_5 q^C) 
     -\frac{\sigma^2}{4 G_S} - \frac{|\Delta|^2}{4 G_D} \, .
\label{mGN}
\end{align}                                                                   
Here $\sigma$ and $\Delta$ are the order parameters for the $q\bar{q}$ and $qq$
condensates, which are defined by
\begin{align}                                                                 
    \sigma = -2G_S \langle \bar{q}q \rangle \,\, , \,\,
    \Delta = 2G_D \langle \bar{q}^C i \tau_2 \lambda_2 \gamma_5 q \rangle.
\end{align}                                                                   

\section{The thermodynamic potential}
The partition function of the ground canonical ensemble is calculated by
using the standard method,
\begin{align}                                                                 
\mathcal{Z} = N^{\prime} \int [d\bar{q}][dq] \exp
  \biggl\{
    \int_0^{\beta}\!\!\! d\tau \! \int \!\! d^2{\bf x} 
       \bigl( \tilde{\mathcal{L}}+\mu \bar{q}\gamma_0 q \bigr)
  \biggr\},
\end{align}                                                                   
where $\mu$ is the quark chemical potential and $\beta=1/T$ is the inverse temperature.
Introducing the Nambu-Gorkov basis\cite{Nambu} 
\begin{align}                                                                 
     \Psi = 
     \left(
     \begin{array}{c}
     q \\
     q^C
     \end{array}\right)
     \quad {\rm and} \quad
     \bar{\Psi} = 
     \left(
     \, \bar{q} \,\,\, \bar{q}^C
     \right) \, ,
\end{align}                                                                   
we have
\begin{align}                                                                 
\mathcal{Z} = & N^{\prime} \exp
  \biggl\{
    - \int_0^{\beta}\!\!\! d\tau \! \int \!\! d^2{\bf x} 
       \biggl( \frac{\sigma^2}{4G_S}+\frac{|\Delta|^2}{4G_D} \biggr)
  \biggr\} \nonumber \\
  & \times 
  \int [d\Psi] \exp
  \biggl\{
    \frac{1}{2} \sum_{n,{\bf p}} \bar{\Psi} \bigl( \beta G^{-1} \bigr) \Psi
  \biggr\}.
\end{align}                                                                   
The matrix $G^{-1}$ is defined by
\begin{align}                                                                 
  G^{-1} =
  & \left(
    \begin{array}{cc}
       (\slash{p} - \sigma + \mu \gamma^0){\bf 1}_f {\bf 1}_c
     & i \tau_2 \lambda_2 \gamma_5 \Delta \\
       i \tau_2 \lambda_2 \gamma_5 \Delta^{*}
       & (\slash{p} - \sigma - \mu \gamma^0){\bf 1}_f {\bf 1}_c
    \end{array}\right),
\end{align}                                                                   
where ${\bf 1}_f$ and ${\bf 1}_c$ are the unit matrix in flavor and color
spaces, respectively. One can compute the thermodynamic potential
$\Omega = -\ln \mathcal{Z} / \beta V$ by following the same procedure as in
\cite{Mei}, 
\begin{align}                                                                 
     \Omega(\sigma,|\Delta|) &= \Omega_{0}(\sigma,|\Delta|)
            + \Omega_{T}(\sigma,|\Delta|), \label{Thermo3D} \\
     \Omega_0  (\sigma,|\Delta|) &= 
      \frac{\sigma^2}{4G_S} + \frac{|\Delta|^2}{4G_D}
      -4  \int \!\! \frac{d^2 p}{(2\pi)^2}
     \bigl[
     E + E_{\Delta}^+ + E_{\Delta}^-
     \bigr], \label{Thermo3D0} \\
     \Omega_T  (\sigma,|\Delta|) &= 
      -4 T \sum_{\pm} \int \!\! \frac{d^2 p}{(2\pi)^2}
     \biggl[
        \ln (1+e^{-\beta E^{\pm}}) + 2 \ln (1+e^{-\beta E_{\Delta}^{\pm}})
     \biggr]. \label{Thermo3DT}
\end{align}                                                                   
where $V$ is the volume of the system and
\begin{align}                                                                 
 & E_{\Delta}^\pm{}^2 \equiv (E \pm \mu)^2 + |\Delta|^2 \,, \,\,
   E^\pm \equiv E \pm \mu\,, \nonumber \\
 & E \equiv \sqrt{\vec{p}^{\,\, 2} + \sigma^2} \, , \,\,
   \vec{p}^{\,\, 2} = p_1^2 + p_2^2\,.
\label{energy}
\end{align}                                                                   
$\Omega_0$ is $T$ independent contribution and
$\Omega_T$ is the $T$ dependent part. $\Omega_0$ is ultraviolet divergent
and we carry out the renormalization through introducing the counter Lagrangian 
density\cite{Kohyama}, which is of the form 
$\mathcal{L}_C = -Z_S \sigma^2/2 -Z_D|\Delta|^2$.
In the present case, the renormalization factors $Z_S$ and $Z_D$ are given by
\begin{align}                                                                 
  Z_S &= \frac{12}{\pi} \Lambda - \frac{3}{2} \alpha \,, \\
  Z_D &= \frac{4}{\pi} \Lambda - \frac{1}{2}\alpha \,.
\end{align}                                                                   
Here $\Lambda$ is the 3D momentum cut-off and $\alpha$ is an arbitrary
renormalization scale.

Then, by introducing the following parameters,
\begin{align}                                                                 
\sigma_0 &\equiv -\frac{\pi}{3}\Bigl( \frac{1}{4G_S} - \frac{3}{4}\alpha \Bigr),
\label{sigma0}\\
\Delta_0 &\equiv -\frac{\pi}{2} \Bigl( \frac{1}{4G_D} - \frac{1}{2}\alpha \Bigr),
\label{delta0}
\end{align}                                                                   
we obtain the renormalized thermodynamic potential:
\begin{align}                                                                 
    \Omega_{r}&(\sigma,|\Delta|) = \Omega_{0r}(\sigma,|\Delta|)
      + \Omega_T(\sigma,|\Delta|)  \,, \label{reThermo} \\
     \Omega_{0r} & (\sigma,|\Delta|) = 
      -\frac{3\sigma_0}{\pi} \sigma^2
      -\frac{2 \Delta_0}{\pi} |\Delta|^2 
      +\frac{2}{3\pi}\sigma^3 \nonumber \\
     & {} + \frac{1}{3\pi} \sum_{\pm} 
     \biggl[
         (2\sigma^2 + 2|\Delta|^2 - \mu^2 \pm \mu \sigma)
         \sqrt{\sigma^2 + |\Delta|^2 + \mu^2 \pm 2\mu\sigma} \nonumber \\
         & \mp 3\mu |\Delta|^2
         \ln \bigl\{
           \sigma \pm \mu + \sqrt{\sigma^2 + |\Delta|^2 + \mu^2 \pm 2 \mu \sigma}
         \bigr\}
     \biggr].
\label{rezero}
\end{align}                                                                   
$\sigma_0$ ($\Delta_0$) is the $q\bar{q}$ ($qq$) condensate at $T=0$ and $\mu=0$
in the model when the $qq$ ($q\bar{q}$) condensate is absent.
It should be noted that the thermodynamic potential has two
free parameters $\sigma_0$ and $\Delta_0$. As in \cite{Kohyama}, we take $\sigma_0$
to be the scale of the theory ($\sigma_0 \geq 0$). Then, after fixing $\sigma_0$,
there remains one free
parameter $\Delta_0$ and we study the various values for the ratio $\Delta_0/\sigma_0$.

\section{Quark-antiquark and diquark condensates}
In this section, we show the numerical result of the $q\bar{q}$ and $qq$ condensates
through analyzing the thermodynamic potential, Eq.(\ref{reThermo}).
The realized condensates are obtained by finding the minimum of the
thermodynamic potential with respect to $\sigma$ and $\Delta$.

\begin{figure*}
\begin{center}
\includegraphics[width=16cm]{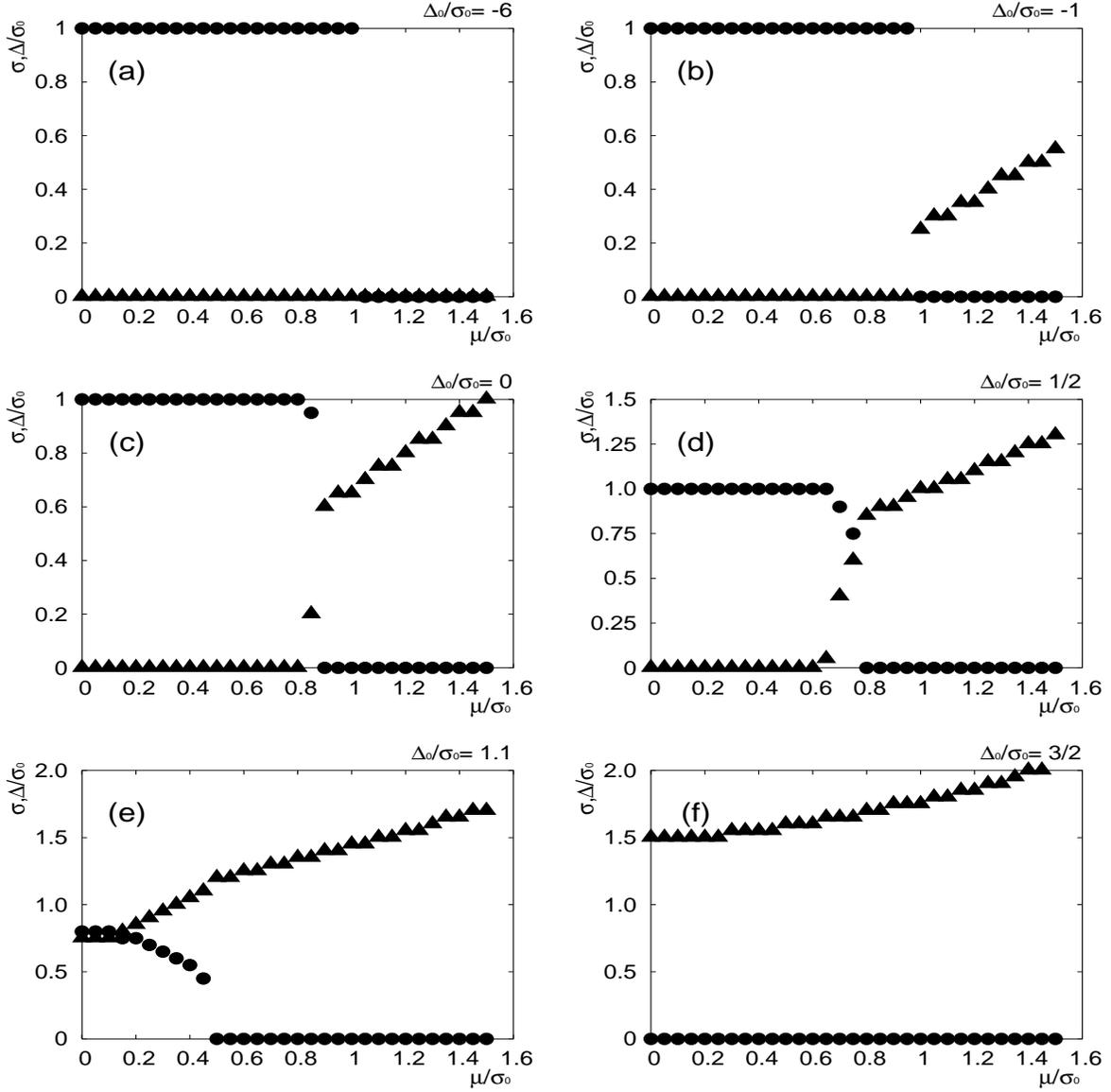}
\end{center}
\caption{\label{fig:0condensate}$\sigma$ (circles) and $\Delta$ 
  (triangles)
  as a function of chemical potential $\mu$ at $T=0$.}
\end{figure*}
Fig.~\ref{fig:0condensate} displays the results of the two condensates 
against $\mu$ at $T=0$.
From the panel (a) ($\Delta_0/\sigma_0=-$6), we see that the $q\bar{q}$ condensate
exists for small $\mu$ and it disappears when $\mu$ becomes large. This clearly shows the
phenomena of the phase transition, and the transition chemical potential is 
$\mu=1.0\sigma_0$. There does not occur the $qq$
condensate for whole $\mu$. However, for $\Delta_0/\sigma_0=-1$, the $qq$ condensate appears 
at $\mu=1.0\sigma_0$, the value where the $q\bar{q}$
condensate disappears. Similar results are obtained in the cases $\Delta_0/\sigma_0 = 0, 1/2$.
At some chemical potential, the $q\bar{q}$ condensate falls and the $qq$ condensate
arises. The transition chemical potentials are $\mu = 0.9\sigma_0$ and $0.8 \sigma_0$
for $\Delta_0/\sigma_0 = 0$ and $1/2$, respectively. Thus the
transition chemical potential becomes smaller with increasing the ratio $\Delta_0/\sigma_0$,
and the value of the $qq$ condensate $\Delta$ enlarges. For $\Delta_0/\sigma_0=1.1$, the $qq$ condensate
at $\mu =0$ has the close value with the $q\bar{q}$ condensate, and it eventually
exceeds the $q\bar{q}$ condensate for $\Delta_0/\sigma_0=3/2$. Note that the
$q\bar{q}$ condensate does not exist for $\Delta_0/\sigma_0=3/2$ and there
appears only the $qq$ condensate. Thus we see that the behaviors of the
$q\bar{q}$ and $qq$ condensates are sensitive to the value of the ratio $\Delta_0/\sigma_0$.

\begin{figure*}
\begin{center}
\includegraphics[width=16cm]{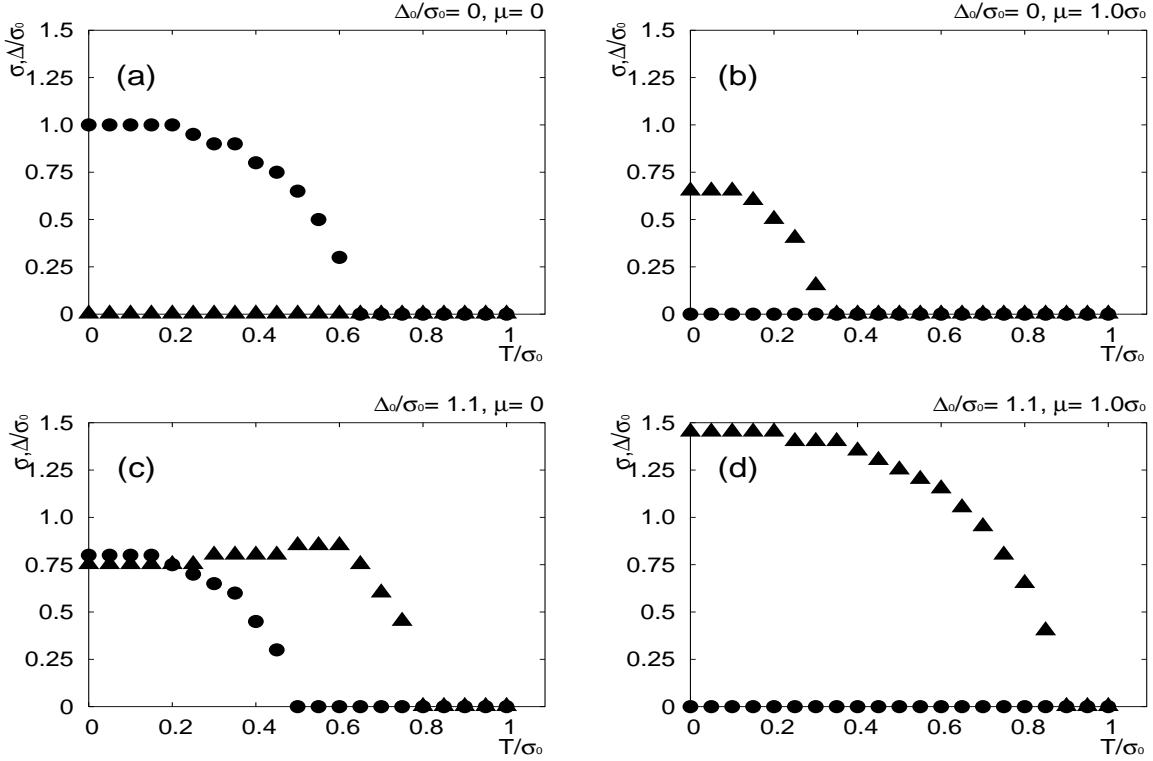}
\end{center}
\caption{\label{fig:Tcondensate}The two gaps $\sigma$ (circles) and $\Delta$ 
  (triangles) against T.}
\end{figure*}
The results of the condensates at finite temperature are shown in Fig.~\ref{fig:Tcondensate}.
In the panel (a) ($\Delta_0/\sigma_0=0$, $\mu = 0$), we see that
the $q\bar{q}$ condensate occurs at $T=0$ and it decreases when
$T$ becomes larger. The transition temperature in this case is $T=0.65 \sigma_0$.
The panel (b) ($\Delta_0/\sigma_0=0$, $\mu = 1.0 \sigma_0$)
shows that the $qq$ condensate is $0.65 \sigma_0$ at $T=0$ and it disappears at $T = 0.35 \sigma_0$.
Thus we see that the $q\bar{q}$ and $qq$ condensates decreases as the temperature increases.
This is also the case for $\Delta_0/\sigma_0=1.1$, which is shown in the panels (c) and (d).
We have analyzed the cases for other values of $\Delta_0/\sigma_0$, and found that the
behavior of the two condensates do not change qualitatively.

It should be noted that, in Fig.~\ref{fig:0condensate} ($T=0$ case), the $q\bar{q}$ condensate
disappears discontinuously. This is the signal of the
first order phase transition. On the other hand, the $q\bar{q}$ condensate disappears
continuously in Fig.~\ref{fig:Tcondensate}(a) ($\mu = 0$ case), which indicates the
second order phase transition.
This means that the critical point from the first order phase transition to the second order
one appears at some point in the phase diagram, which we will discuss in more detail in
the next section.

\section{The phase diagram}
\begin{figure*}
\begin{center}
\includegraphics[width=16cm]{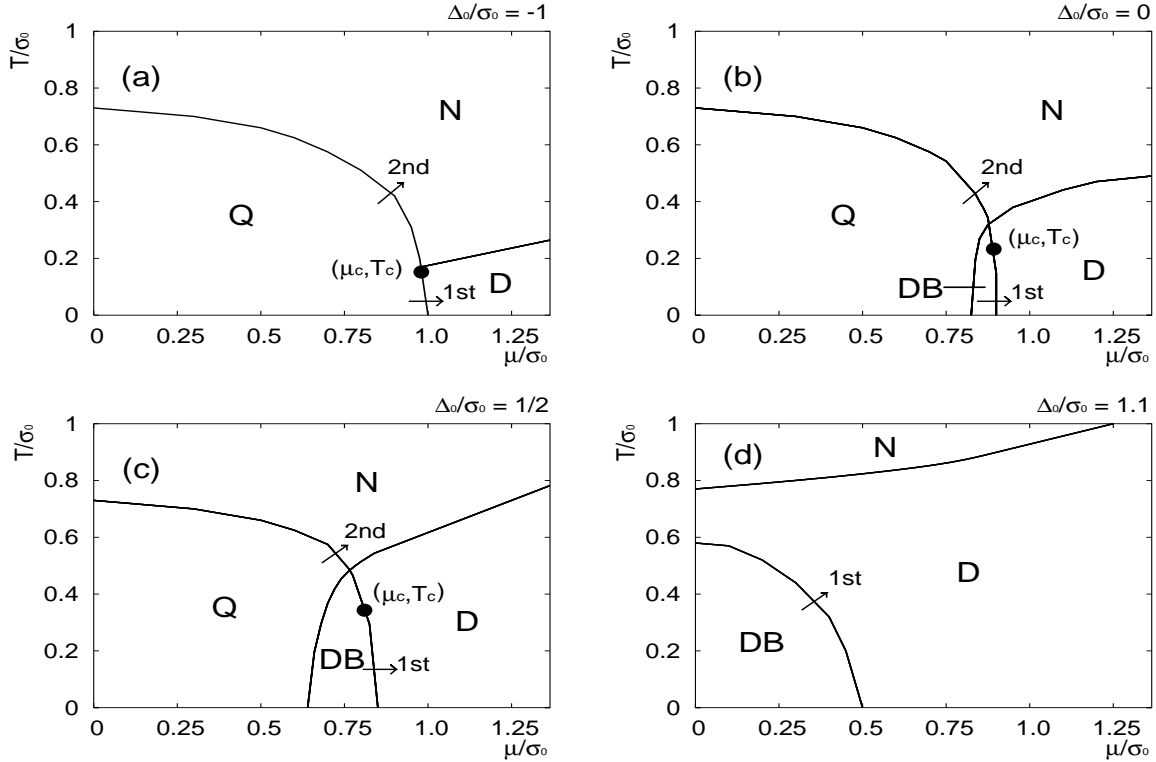}
\end{center}
\caption{\label{fig:phase}The phase diagrams.}
\end{figure*}

On the basis of the results obtained in the previous section, we construct the phase diagram.
Fig.~\ref{fig:phase} displays the phase structure of the $q\bar{q}$
and $qq$ condensates for the cases 
$\Delta_0/\sigma_0= -1,\, 0,\, 1/2,\, 1.1$.

From the panel (a), we see that the $q\bar{q}$ condensate phase is realized at low $T$
and $\mu$, and the $qq$ condensate phase occurs at low $T$ and high $\mu$. At high
$T$ and $\mu$, there does not exist condensate, namely the system is in the normal phase.
The similar structure is seen in the panel (b):
$q\bar{q}$ phase at low $T$ and $\mu$, $qq$ phase at low $T$ and high $\mu$. However,
in the case of $\Delta_0/\sigma_0=0$,
we see that the coexisting phase of the $q\bar{q}$ and $qq$ condensates appears at
low $T$ and intermediate $\mu$.
We refer this phase to as the ``Double Broken" (DB) phase. Thus, the system is
characterized by the following phases:
\begin{quote}
 Q : $q\bar{q}$ condensate phase ($\sigma \neq 0$,$\,\,\Delta = 0$)\\
 D : $qq$ condensate phase ($\sigma = 0$,$\,\,\Delta \neq 0$)\\
 DB : Double broken phase ($\sigma \neq 0$,$\,\,\Delta \neq 0$)\\
 N : Normal phase ($\sigma = 0$,$\,\,\Delta = 0$)
\end{quote}
All of these phases appear in the case of $\Delta_0/\sigma_0=1/2$ (panel (c)). However the
Q phase does not appear for $\Delta_0/\sigma_0=1.1$ and only D phase and DB phase
appear. Thus, as the ratio $\Delta_0/\sigma_0$ increases,
the Q condensate phase shrinks toward $\mu$ axis and the region of D phase and DB phase
become larger.

The points ($\mu_c,T_c$) in the panels (a), (b) and (c) indicate the critical points
from the first order phase transition to the second order one. The phase transition below the
critical temperature $T_c$ is of the first order and above $T_c$ is of the
second order. We notice that the value of $T_c$ increases when $\Delta_0/\sigma_0$ becomes
larger, while $\mu_c$ decreases with increasing $\Delta_0/\sigma_0$.
Then the critical point moves upward along the transition line, and the region of the
first order phase transition expands.
The other phase transitions, namely the transitions Q $\rightarrow$ DB,
Q $\rightarrow$ N and D $\rightarrow$ N, are of the second order.

\section{Concluding remarks}
Through studying the $q\bar{q}$ and $qq$ condensates at finite temperature and
chemical potential, we have obtained the phase diagrams in the 3D GN model
with the 4c quarks.

We have shown that the $q\bar{q}$ condensate (Q) phase is realized at low $T$ and $\mu$,
the double broken (DB) phase at low $T$
and intermediate $\mu$, and the $qq$ condensate (D) phase at low $T$ and high $\mu$
(see Fig.~\ref{fig:phase}(b) and (c)). This feature bears resemblance
to the case of the 4D NJL model. It is difficult to make a direct comparison between the
present model and the 4D NJL model, because the free parameter in the NJL model is
the ``direct ratio" of the coupling constants $G_D/G_S$, while the free parameter
of the present model is the ratio $\Delta_0/\sigma_0$ which is not $G_D/G_S$.
However the parameter $\Delta_0$ is related to $G_D$ through Eq.(\ref{delta0}),
and $\Delta_0$ becomes larger as $G_D$ increases. Then the ratio
$\Delta_0/\sigma_0$ increases as the $qq$ coupling constant $G_D$ increases.
This means that, as $G_D/G_S$ increases, the region of the Q phase shrinks,
and the regions of the D phase and the DB phase
expand. Then, the behavior of the phase diagrams shows the
close similarity to the case of the 4D NJL model.

We are now in the position to make the comparison with the 3D GN model with the
2c quarks. Comparing with the phase diagrams in the 2c quark case obtained in \cite{Kohyama},
we see rather similar structure: The Q phase at low $T$ and $\mu$, and the D phase
at low $T$ and high $\mu$. However, there is one crucial difference.
There does not appear the DB phase in the 2c quark case,
while in the present 4c quark case, the DB phase does appear. The latter fact is in
accord with the expectation mentioned in the introduction.

The critical points between the first-order and second-order transition,
are located on the Q $\rightarrow$ D
phase transition line in Fig.~\ref{fig:phase}(a) and on the DB $\rightarrow$ D transition
line in Fig.~\ref{fig:phase}(b) and (c). In the model without $qq$ condensate,
the $q\bar{q}$ phase transition at zero temperature is of the first order and the transition
at finite temperature is of the second order. This means that $T_c$ is negligibly
small in the model without the $qq$ condensate. While in the present model, $T_c$ is finite,
which is the reflection of the existence of the
$qq$ condensate. It is also worth comparing the present model with the model with the
2c quarks, where the coexisting phase is absent.
The phase transition Q $\rightarrow$ D is always of the first order and the critical
points are seen on the line Q $\rightarrow$ N\cite{Kohyama}.
This may come from
the fact that the existence of the $qq$ condensate expels the $q\bar{q}$ condensate.
On the other hand, in the present 4c quark case, the $q\bar{q}$ and $qq$ condensates
can coexist and the Q $\rightarrow$ D and DB $\rightarrow$ D transitions can be of
the second order.

Finally, it is worth reemphasizing that the significant qualitative difference is
that there exists the $q\bar{q}$ and $qq$ coexisting phase in the present model, which is not
seen in the 2c quark case. Then, when compared to the case of the 3D GN model with the 2c quarks,
the phase structures of the present 4c quark case bear closer resemblance to the 4D NJL model.

\begin{acknowledgments}
I would like to express my sincere gratitude to A. Niegawa for useful discussions
and reading of the manuscript. The valuable discussions with M. Inui and T. Inagaki
are also gratefully acknowledged.
\end{acknowledgments}




\end{document}